\documentclass[%
 aps,
 prapplied,
 superscriptaddress,
 longbibliography,
 amsmath,amssymb,
 reprint,%
]{revtex4-1}

\usepackage{ mathrsfs  }
\usepackage{graphicx}
\usepackage{dcolumn}
\usepackage{bm}
\usepackage[dvipsnames]{xcolor}

\usepackage{gensymb}

\newcommand{\gpc}{\dot\gamma_{\rm c}}
\newcommand{\gpp}{\dot\gamma_{\rm p}}
\newcommand{\gp}{\dot\gamma}

\definecolor{colordg}{rgb}{0,0.65,0.45}
\definecolor{colorlg}{rgb}{0,0.9,0.65}
\definecolor{colorlb}{rgb}{0,0.65,0.95}
\definecolor{colorb1}{rgb}{0,0.9255,0.5373}
\definecolor{colorb2}{rgb}{0,0.7294,0.6353}
\definecolor{colorb3}{rgb}{0,0.4549,0.7725}
\definecolor{colorb4}{rgb}{0,0.1412,0.9294}
\definecolor{colorg1}{rgb}{0,0.8,0.6}
\definecolor{colorg2}{rgb}{0,0.6,0.6}
\definecolor{colorgp2}{rgb}{1,0.4,0.1}
\definecolor{colorgp5}{rgb}{1,0.2,0.2}
\definecolor{colorgp10}{rgb}{1,0.5,0.5}
\definecolor{colorgp100}{rgb}{0.2,0.5,1}
\definecolor{colorgp1000}{rgb}{0.1,0.7,1}
\definecolor{colorX1}{rgb}{0,0,0}
\definecolor{colorX2}{rgb}{0.6,0.6,0.6}

\begin{document}

\title[Rheo-SAXS study of reinforcement in boehmite gels]{Shear-induced reinforcement in boehmite gels:\\a rheo-X-ray-scattering study
}

\author{Iana Sudreau}
\affiliation{IFP Energies Nouvelles, Rond-point de l'{\'e}changeur de Solaize, BP 3, 69360 Solaize, France}
\affiliation{ENSL, CNRS, Laboratoire de physique, F-69342 Lyon, France}
\author{Marion Servel}
\affiliation{IFP Energies Nouvelles, Rond-point de l'{\'e}changeur de Solaize, BP 3, 69360 Solaize, France}
\author{Eric Freyssingeas}
\affiliation{ENSL, CNRS, Laboratoire de physique, F-69342 Lyon, France}
\author{Fran{\c{c}}ois Li{\'e}nard}
\affiliation{ENSL, CNRS, Laboratoire de physique, F-69342 Lyon, France}
\author{Szilvia Karpati}
\affiliation{Universit{\'e} Lyon 1, Ecole Normale Sup{\'e}rieure de Lyon, CNRS, UMR 5182, Laboratoire de Chimie, 46 all{\'e}e d'Italie, Lyon F69364, France}
\author{St{\'e}phane Parola}
\affiliation{Universit{\'e} Lyon 1, Ecole Normale Sup{\'e}rieure de Lyon, CNRS, UMR 5182, Laboratoire de Chimie, 46 all{\'e}e d'Italie, Lyon F69364, France}
\author{Xavier Jaurand}
\affiliation{Univ Lyon, Universit{\'e} Lyon 1, CT$\mu$, F-69622, Villeurbanne Cedex, France}
\author{Pierre-Yves Dugas}
\affiliation{ Univ Lyon, Universit{\'e} Lyon 1, CPE Lyon, CNRS, UMR 5128, Catalysis, Polymerization, Processes and Materials (CP2M), F-69616 Villeurbanne, France}
\author{Lauren Matthews}
\affiliation{ESRF - The European Synchrotron, 38043 Grenoble Cedex, France }
\author{Thomas Gibaud}
\affiliation{ENSL, CNRS, Laboratoire de physique, F-69342 Lyon, France}
\author{Thibaut Divoux}
\affiliation{ENSL, CNRS, Laboratoire de physique, F-69342 Lyon, France}
\author{S{\'e}bastien Manneville}
\affiliation{ENSL, CNRS, Laboratoire de physique, F-69342 Lyon, France}
\affiliation{Institut Universitaire de France (IUF)}

\date{\today}

\begin{abstract}
Boehmite, an aluminum oxide hydroxide $\gamma$-AlO(OH), is broadly used in the form of particulate dispersions in industrial applications, e.g., for the fabrication of ceramics and catalyst supports or as a binder for extrusion processes. Under acidic conditions, colloidal boehmite dispersions at rest form gels, i.e., space-spanning percolated networks that behave as soft solids at rest, and yet yield and flow like liquids under large enough deformations. Like many other colloidal gels, the solid-like properties of boehmite gels at rest are very sensitive to their previous mechanical history. Our recent work [Sudreau \textit{et al.}, J. Rheol. 66, 91-104 (2022), and Phys. Rev. Material 6, L042601 (2022)] has revealed such \textit{memory effects}, where the shear experienced prior to flow cessation drives the elasticity of boehmite gels: while gels formed following application of a shear rate $\gpp$ larger than a critical value $\gpc$ are insensitive to shear history, gels formed after application of $\gpp<\gpc$ display reinforced viscoelastic properties and non-negligible residual stresses. Here, we provide a microstructural scenario for these striking observations by coupling rheometry and small-angle X-ray scattering. Time-resolved measurements for $\gpp <\gpc$ show that scattering patterns develop an anisotropic shape that persists upon flow cessation, whereas gels exposed to $\gpp>\gpc$ display isotropic scattering patterns upon flow cessation. Moreover, as the shear rate applied prior to flow cessation is decreased below $\gpc$, the level of anisotropy frozen in the sample microstructure grows similarly to the viscoelastic properties, thus providing a direct link between mechanical reinforcement and flow-induced microstructural anisotropy.
\end{abstract}

\maketitle

\section{Introduction}

Suspensions of attractive particles are encountered in a broad range of natural and engineered systems, from foodstuff, and natural clays to semi-solid electrodes and cementitious materials \cite{DelGado:2014,Suman:2018,Narayanan:2021,Cao:2020}. Such suspensions flow like liquids under large external shear, whereas they behave like soft solids upon flow cessation \cite{Bonn:2017}. Indeed, attractive interparticle forces drive the formation of a space-spanning network, which morphology depends on the details of diffusive transport during the sol-gel transition \cite{Johnson:2019,Gibaud:2022}. On the one hand, strong attraction leads to the rapid formation of a microstructure composed of open clusters that result in a network of string-like strands. On the other hand, weak attractive interactions favor a phase separation scenario \cite{Zaccarelli:2007,Zaccarelli:2008}, leading to a network of compact, thicker strands with a glassy-like behavior \cite{Zaccone:2009,Whitaker:2019,Koumakis:2011,Zia:2014}. 

Irrespective of the gelation scenario, it has long been recognized that the mechanical history of attractive colloidal dispersions prior to gelation at rest could strongly impact their subsequent viscoelastic behavior \cite{Lindstrom:2012,Ovarlez:2013,Koumakis:2015}. In order to optimize the processing of colloidal gels, it is thus essential to understand the interplay between the shear history and the rheological properties, as well as the final microstructure obtained after flow cessation. Indeed, during the liquid-to-solid transition following flow cessation, structural anisotropy that develops under shear may remain ``frozen'' in the gel microstructure, thus impacting its macroscopic mechanical properties. This picture has been strongly supported by experimental measurements based on a variety of techniques including rheo-electrical spectroscopy, orthogonal superposition rheometry, and small-angle X-ray scattering under shear (rheo-SAXS) \cite{Varadan:2001,Helal:2016,Colombo:2017,Narayanan:2017,Dages:2022}, and confirmed by numerical simulations \cite{Moghimi:2017}. Moreover, the structural anisotropy encoded into the gel microstructure may also affect the non-linear mechanical response, resulting in kinematic hardening \cite{Larson:2019}. For instance, stress-induced failure of a colloidal gel occurs at lower strain when a creep test is performed in the opposite direction to that of the shear applied prior to gelation at rest \cite{Grenard:2014}. Similar observations during flow reversal experiments under applied shear rate on fumed silica suspensions show that the yield stress is increased along the shear direction, akin to the so-called Bauschinger effect also observed in denser soft glassy materials \cite{Wei:2019,Kushnir:2022}. 

Building upon the idea of encoding some ``memory'' of previous mechanical history into a colloidal gel, shear was used as a way to finely tune the structural and mechanical properties of colloidal gels. In particular, it was shown that the elasticity of depletion gels is reinforced after applying a large shear rate, which yields more homogeneous, highly connected gels \cite{Koumakis:2015}. In contrast, carbon black gels and boehmite gels, in which the driving attractive interparticle force originates from van der Waals interactions, display significant shear-induced reinforcement following the application of low shear rates \cite{Sudreau:2022a,Dages:2022}. In the case of carbon black gels, the reinforcement was attributed to the degree of interpenetration of fractal clusters, which increases for decreasing shear rates \cite{Dages:2022}, or to the formation of clusters of clusters \cite{Bouthier:2023}. As for boehmite gels, Sudreau {\it et al.} \cite{Sudreau:2022a} hypothesized that the enhancement of elasticity results from the build-up of some structural anisotropy within the colloidal clusters, a scenario that is supported by unexpected non-monotonic stress relaxations following cessation of flow at low shear rates \cite{Sudreau:2022b}. Finally, some gels of attractive spheres were shown to exhibit strengthening, while attractive rod gels display weakening within the same range of shear rates, pointing to a strong impact of particle shape on gel structuring under flow \cite{Das:2022}.

The above observations suggest that, to date, there is no universal mechanism accounting for shear-induced reinforcement in colloidal gels and that more structural and rheological experimental data are required to elucidate the subtle interplay between shear history, microstructural anisotropy, and viscoelastic properties. The present article aims to unveil such an interplay in the specific case of boehmite gels.
To this aim, we focus on acid-induced gels of microcrystalline boehmite ($\gamma$-AlOOH) and perform time-resolved SAXS experiments both under shear and after flow cessation. 
Our results unravel the growth of anisotropy under external shear, which remains frozen within the gel microstructure for applied shear rates $\gpp$ smaller than a critical value $\gpc$, which  corresponds to the onset of elastic reinforcement. Moreover, the level of anisotropy encoded upon flow cessation increases for decreasing values of $\gpp$, in the same manner as the elastic modulus $G'$. This finding constitutes the core of the present article, which outline goes as follows.
After summarizing our previous work in Sect.~\ref{sec:summary},
we describe the various techniques used to characterize the microstructure of boehmite dispersions in Sect.~\ref{sec:materials}. Section~\ref{sec:microstructure} shows that the building blocks in boehmite gels are constituted of polydisperse, fractal-like aggregates of individual crystallites. Section~\ref{sec:results} then provides evidence for shear-induced frozen-in anisotropy and for its correlation with the enhancement of elasticity. Finally, Sect.~\ref{sec:discussion} offers a short discussion and conclusion.

\section{Summary of previous work}
\label{sec:summary}

Boehmite gels consist in dispersions of colloidal anisotropic particles, whose exact shape depends on the synthesis \cite{Music:1999,Chen:2007}. They serve as soft precursors for alumina-derived materials of controlled porosity, such as catalyst supports \cite{Trimm:1986,Euzen:2002,Zheng:2014}, and binders for alumina extrusion \cite{Ananthakumar:2001}. The gel microstructure is governed by attractive interactions between the particles, which can be tuned by playing on the ionic strength of the dispersion \cite{Wierenga:1998}. Over a broad range of salt content, boehmite gels display a fractal-like microstructure, while their linear viscoelastic properties increase as a power law of the particle volume fraction \cite{Shih:1990}. From a rheological perspective, boehmite gels behave as 
yield stress fluids with thixotropic properties
\cite{Ramsay:1978,Sudreau:2022a}.

In two recent experimental contributions \cite{Sudreau:2022a,Sudreau:2022b}, we have characterized the mechanical properties of 4\%~vol. boehmite gels obtained upon flow cessation (see Sect.~\ref{sec:sample_prep} below for the detailed preparation protocol). In brief, a gel sample was loaded into a Taylor-Couette shear cell and first rejuvenated by a constant shear
rate $\gp=1000$~s$^{-1}$ during 600~s. The flow was then quenched at a given shear rate $\gpp$ during 600~s. Upon cessation of such a shear (induced by applying $\gp=0$), the viscoelastic properties of the fluidized gel were monitored under small-amplitude oscillatory shear (with frequency $f=1$~Hz and amplitude $\gamma=0.1$\%) during 3000~s. The elastic modulus $G'$ was observed to progressively increase above the viscous modulus $G''$. The terminal value of the elastic modulus $G'$ was recorded and the experiment was repeated for various values of $\gpp$ ranging between 2~s$^{-1}$ and 1000~s$^{-1}$. \begin{figure}[t!]
    \centering
    \includegraphics[width=1\linewidth]{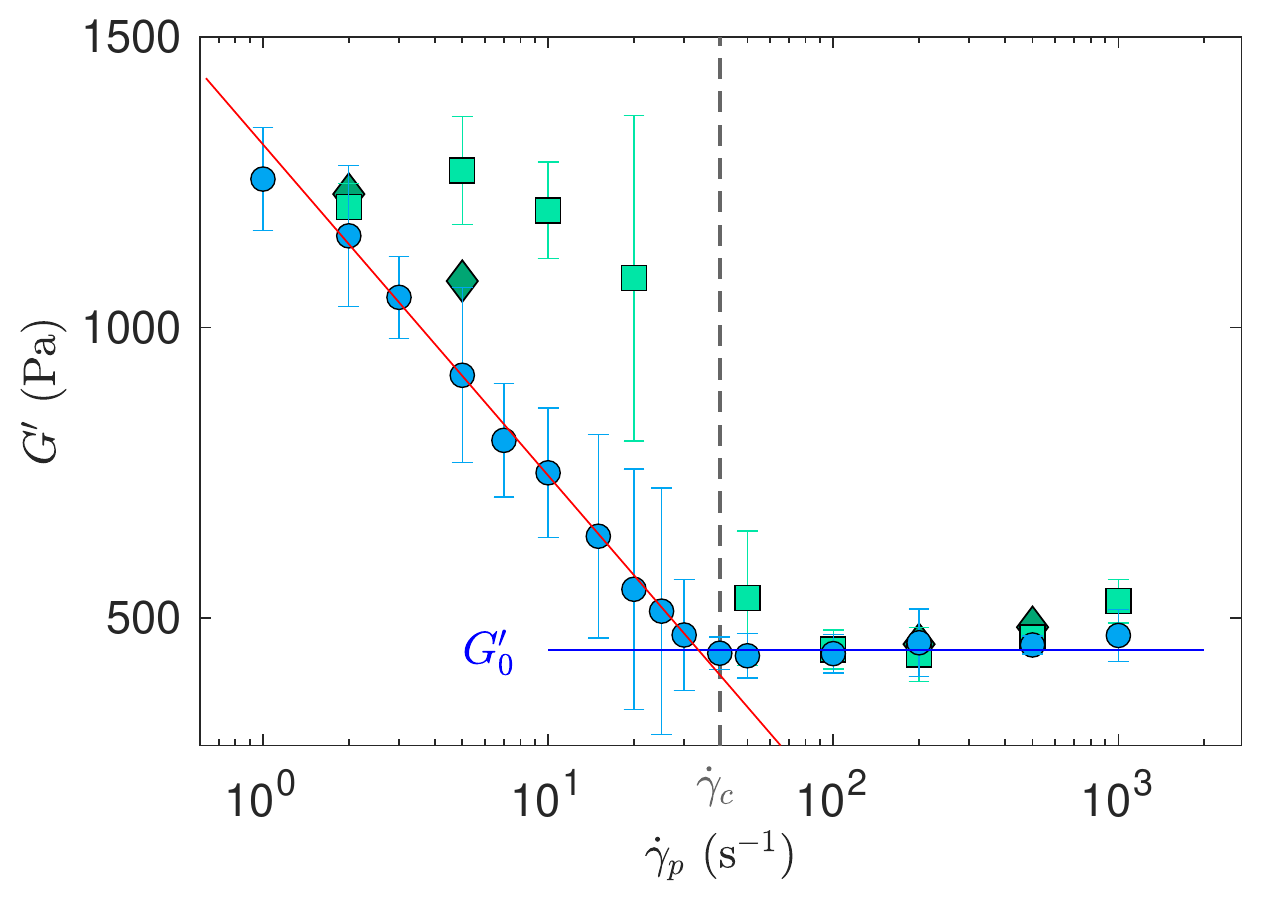}
    \caption{Elastic modulus $G'$ as a function of the shear rate $\dot \gamma_p$ applied prior to flow cessation. $G'$ is measured at $t=3000$~s after flow cessation for protocols A (\textcolor{colorlb}{$\bullet$}), B (\textcolor{colorlg}{$\blacksquare$}), and C (\textcolor{colordg}{$\blacklozenge$}). The vertical dashed line highlights the critical shear rate $\gpc\simeq 40$~s$^{-1}$ below which $G'$ is reinforced. The blue horizontal line is $G'_{0} = 440~\pm 35$~Pa, and the red line is the best logarithmic fit of the data for $\gpp<\gpc$: $G'=a_{G'}\log\gpp+b_{G'}$ with $a_{G'}=-570$ Pa and $b_{G'}=1 315$ Pa. Error bars correspond to the standard deviations computed over three to six independent measurements.}
    \label{fig:robustesse_protocole}
\end{figure}
Such a rheological protocol, referred to here as protocol ``A" and introduced in Ref.~\cite{Sudreau:2022b}, leads to the dependence of the terminal value of $G'$ with $\gpp$ reported with blue dots in Figure~\ref{fig:robustesse_protocole} .

Remarkably, a critical shear rate $\gpc\simeq 40$~s$^{-1}$ separates two radically different behaviors: gels obtained following strong shear, i.e., $\gpp>\gpc$, are insensitive to $\gpp$ and display a constant elastic modulus $G'=G'_0\simeq 440$~Pa. In contrast, for gels formed following weaker shear, i.e., $\gpp<\gpc$, $G'$ increases logarithmically for decreasing values of $\gpp$. 
In order to test for the robustness of such a striking reinforcement of elasticity, we have tested two additional shear protocols. In protocol ``B" introduced in Ref.~\cite{Sudreau:2022a}, the rejuvenation step at $\gp=1000$~s$^{-1}$ in between two successive values of $\gpp$ was suppressed, the rest period was  followed by a frequency sweep and a strain sweep. In  protocol ``C", we performed the same series of steps as in protocol ``A", yet with an  additional period of rest period for 3000~s after the 600~s period of shear at $\gp=1000$~s$^{-1}$. This allows to check that the effects obtained starting directly from a fully fluidized state (protocol ``A'') are also present when starting from a solid-like state (protocol ``C'') or from partially rejuvenated states (protocol ``B''). As illustrated in Fig.~\ref{fig:robustesse_protocole}, in spite of a more abrupt transition with protocol ``B'' than with protocols ``A'' and ``C'', all three protocols point to the same value of $\gpc$. This demonstrates the robustness of the critical shear rate below which the gel shows strongly reinforced elasticity. In the following, we proceed with a microstructural characterization of boehmite gels, in order to uncover the microscopic origin of the effect of $\gpp$ shown in Fig.~\ref{fig:robustesse_protocole} on the gel final elasticity.

\section{Materials and methods}

\label{sec:materials}

\subsection{Preparation of boehmite gels}
\label{sec:sample_prep}

Colloidal suspensions of boehmite were prepared from Pural SB3 powder produced by Sasol via the hydrolysis of aluminum  alcoholate (Ziegler process) \cite{Wang:2010}. 
The powder was dispersed in an aqueous solution of nitric acid. The mixture was stirred at 600~rpm during 20 min, and then at 1800~rpm during 15~min (Turrax IKA RW20 equipped with dissolver stirrers). The samples were prepared at a reference concentration of 123~g.L$^{-1}$ in boehmite, which corresponds to a solid volume fraction of 4$\%$ vol., and 14~g.L$^{-1}$ in nitric acid \cite{Sudreau:2022a}. 
A sol--gel transition occurs within the first couple of hours, while the  stabilization of the pH at $\text{pH}=3.5$ takes several days due to the slow partial dissolution of the boehmite surface, forming polycations that, in turn, adsorb onto the boehmite surface  \cite{Strenge:1991,Cristiani:2007,Fauchadour:2002}. The presence of nitrate anion NO$_{3}^{-}$ screens the repulsive electrostatic interactions between the positively charged surfaces of the boehmite particles, thus triggering the formation of a space-spanning percolated network \cite{Wood:1990,Speyer:2020,Raybaud:2001,Zheng:2014}.

\subsection{SEM images of the boehmite raw powder}

Images of the raw boehmite powder were acquired with a scanning electron microscope (SEM, Zeiss Supra 55 VP). Two working distances, defined as the distance between the last lens of the microscope and the sample, were used: 5.4~mm and 6.9~mm, with accelerating voltages of 3~kV and 4~kV, respectively.

\subsection{TEM images of boehmite raw powder and crystallites}
\label{sec:TEM}

Transmission Electron Microscopy (TEM, JEOL 1400Flash) was further used to image the raw boehmite powder and crystallites. Images were taken under an accelerating voltage of 120~kV with a Gatan Orius 600 camera and processed with Digital Micrograph software.  
On the one hand, dispersions were prepared by suspending 11.9~mg of boehmite powder in 5 mL of ethanol under application of ultrasound at a power of 160~W for 15 min with an ultrasonic bath (Bandelin Sonorex Digitec). The suspension was then left at rest for one day before being diluted by a factor of 10 and sonicated again at 160~W for 15 min. On the other hand, samples of acidified boehmite suspensions were diluted by a factor of 500 in distilled water, either before or after acidification. 
Finally, all samples are manually homogenized and a deposit of 5~$\mu$L of suspension was made on a copper TEM grid. The deposits were left to dry in the open air before being observed.

\subsection{Cryo-TEM images of boehmite crystallites}

Thin slices of frozen boehmite gels were observed by cryo-TEM, which allows one to study the morphology of dispersed particles while avoiding the evaporation phase of the solvent. Experiments were performed under an accelerating voltage of 120~kV with a JEOL 1400Flash TEM equipped with a Gatan RIO 16 Mpx camera at Centre Technologique des Microstructures (CT$\mu$, Universit{\'e} Claude Bernard Lyon 1, Villeurbanne, France).

Two samples were observed, namely a 4\%~vol. boehmite gel and the same sample diluted by a factor of 3. The gel dilution was performed with distilled water no later than 30~s before sample freezing. 4~$\mu$L of each sample was deposited on a Lacey carbon supported copper grid (300 mesh) previously hydrophilized. In order to form an ultra-thin film on the grid, the excess sample was absorbed thanks to a filter paper. The sample was then immersed in liquid ethane at a temperature of -178$^\circ$C with a freezing apparatus (Thermo Fisher Vitrobot). Finally, the frozen samples were transferred to the TEM cryogenic sample holder (Fischione model 2550), which was maintained at a temperature of -172$^\circ$C. A size analysis was performed on some of the micrographs using the Fiji software (see Sect.~\ref{sec:electron} below).

\subsection{Dynamic Light Scattering}
\label{sec:DLSmethod}

Dynamic Light Scattering (DLS) experiments were performed on boehmite gels at various degrees of dilutions as listed in Table~\ref{tab:DLS_eau} using a standard setup (Brookhaven BI-200SM). A solid-state laser of wavelength $\lambda=532$~nm (Gem 532, Laser Quantum) was used as a light source, which power and polarization were tuned by a set of optical elements. A convergent lens focused the beam on the sample contained in a glass cylinder of inner diameter 8~mm immersed in an index-matching bath of decahydronaphthalene, whose temperature was controlled at $20^\circ$C. A goniometer allowed us to vary the angle $\theta$ between the incident beam and the optical fiber detecting the photons scattered by the sample from $\theta=15^\circ$ to 150$^\circ$. This corresponds to a range of scattering wave vectors of magnitude $q=4 \pi n \sin (\theta/2)/\lambda=0.004$~nm$^{-1}$ to 0.03 nm$^{-1}$, with $n=1.33$ the refractive index of water. 

The scattered light was collected by a multi-mode fiber and detected by an avalanche photodiode module (PerkinElmer SPCM-AQRH). 
The latter signal $I(q,t)$ was recorded over 1~min and processed by a correlator (Brookhaven TurboCorr) to compute the autocorrelation function of the scattered intensity $g^{(2)}(q,\tau)=\langle I(q,t)I(q,t+\tau)\rangle$, where $\langle\cdots\rangle$ denotes the average over $t$. The lag time $\tau$ spanned six orders of magnitude from $\tau=1~\mu$s to 1~s. The dynamics of the system are given by the normalized autocorrelation function $g^{(1)}(q,\tau)$ of the scattered electric field $E_s(q,t)$: $g^{(1)}(q,\tau)=\langle E_s^*(q,t)E_s(q,t+\tau)\rangle/<I(q,t)>$. The latter first-order autocorrelation function is related to the measured $g^{(2)}(q,\tau)$ through the Siegert relation, $g^{(2)}(q,\tau)=\langle I(q,t)\rangle^2 (1+\beta\vert g^{(1)}(q,\tau)\vert^2)$, where $0\le\beta\le 1$ is the inverse of the number of coherence areas on the multi-mode fiber.

\subsection{Small Angle X-ray Scattering}

Small Angle X-ray Scattering (SAXS) experiments were performed at the ID02 beamline (ESRF, Grenoble, France) using an X-ray wavelength of 1~\AA. SAXS experiments were carried at two different sample-to-detector distances, 31~m corresponding to the ultra-small angle range, and 3~m. This allowed us to cover a scattering vector $q$ range of 0.001~nm$^{-1}$ to 2~nm$^{-1}$. Measured 2D SAXS patterns were normalized to an absolute scale after applying different corrections on the fly (see Ref.~\cite{Narayanan.2022} for full technical details). On the one hand, static SAXS measurements were performed on gel samples diluted with distilled water and homogenized manually, prior to loading into a borosilicate capillary of diameter~1~mm. Together with DLS, this allowed us to analyze the structure of the building blocks in boehmite gels, as described in Sect.~\ref{sec:microstructure}. 

\begin{table}[t!]
\caption{\label{tab:DLS_eau}Hydrodynamic radius of boehmite gels building blocks as determined by DLS, following three different homogenization processes. The various dilution factors allow us to ensure that building blocks are dispersed as individual aggregates. Error bars for $R_H$ are deduced from the uncertainty on the fit parameters in Eq.~\eqref{eq:fit_gamma_dls_bis} and \eqref{eq:fit_dls}.}
\centering
\begin{tabular}{ccc}
\\
Dilution factor & Homogenization & $R_H$ (nm) \\
\hline
\hline
\\
14 & Manual & 70 $\pm$ 2\\
30 & Manual & 62 $\pm$ 2 \\
217 & Manual & 52 $\pm$ 2\\
643 & Manual & 51 $\pm$ 2\\
\\
30 & Shear & 57 $\pm$ 3\\
217 & Shear & 54 $\pm$ 2\\
643 & Shear & 56 $\pm$ 3\\
\\
14 & Sonication & 21 $\pm$ 1\\
30 & Sonication & 22 $\pm$ 1 \\
217 & Sonication & 25 $\pm$ 1 \\
643 & Sonication & 28 $\pm$ 2 \\
\end{tabular}
\end{table}

On the other hand, SAXS measurements were performed coupled with rheometry in order to characterize the microstructure of boehmite gels under and after flow. Gel samples were loaded into a polycarbonate Taylor-Couette cell composed of a bob of radius 10 mm and a cup of radius 11 mm,  40~mm height) connected to a stress-controlled rheometer (Haake RS6000, Thermo Scientific) \cite{Narayanan.2020}. We used the so-called ``radial'' configuration, where the X-ray beam passes across the Couette cell along the direction of the velocity gradient $\nabla\vec{v}$, i.e., through the rotation axis of the bob. The scattering patterns were recorded into the (velocity $\vec{v}$, vorticity~$\nabla\times \vec{v}$) plane. In order to isolate the relatively weak signal induced by the gel microstructure from the background signal, a measurement of the intensity scattered by the Couette cell filled with distilled water was systematically subtracted from the intensity recorded on boehmite suspensions. This rheo-SAXS setup allowed us to perform time-resolved X-ray scattering measurements while applying a specific shear protocol to the sample under study, as analyzed in details in Sect.~\ref{sec:results}. 

\section{Structure of the building blocks in boehmite gels}
\label{sec:microstructure}

In this Section, we use the different techniques listed in Sect.~\ref{sec:materials} to identify and characterize the building blocks in boehmite gels. We show that such gels are composed of elementary platelet-like particles referred to as crystallites, which assemble into ``unbreakable'' fractal-like aggregates. When left at rest, these aggregates further arrange into a space-spanning network, leading to solid-like properties.

\begin{figure}[t!]
    \centering
    \includegraphics[width=1\linewidth]{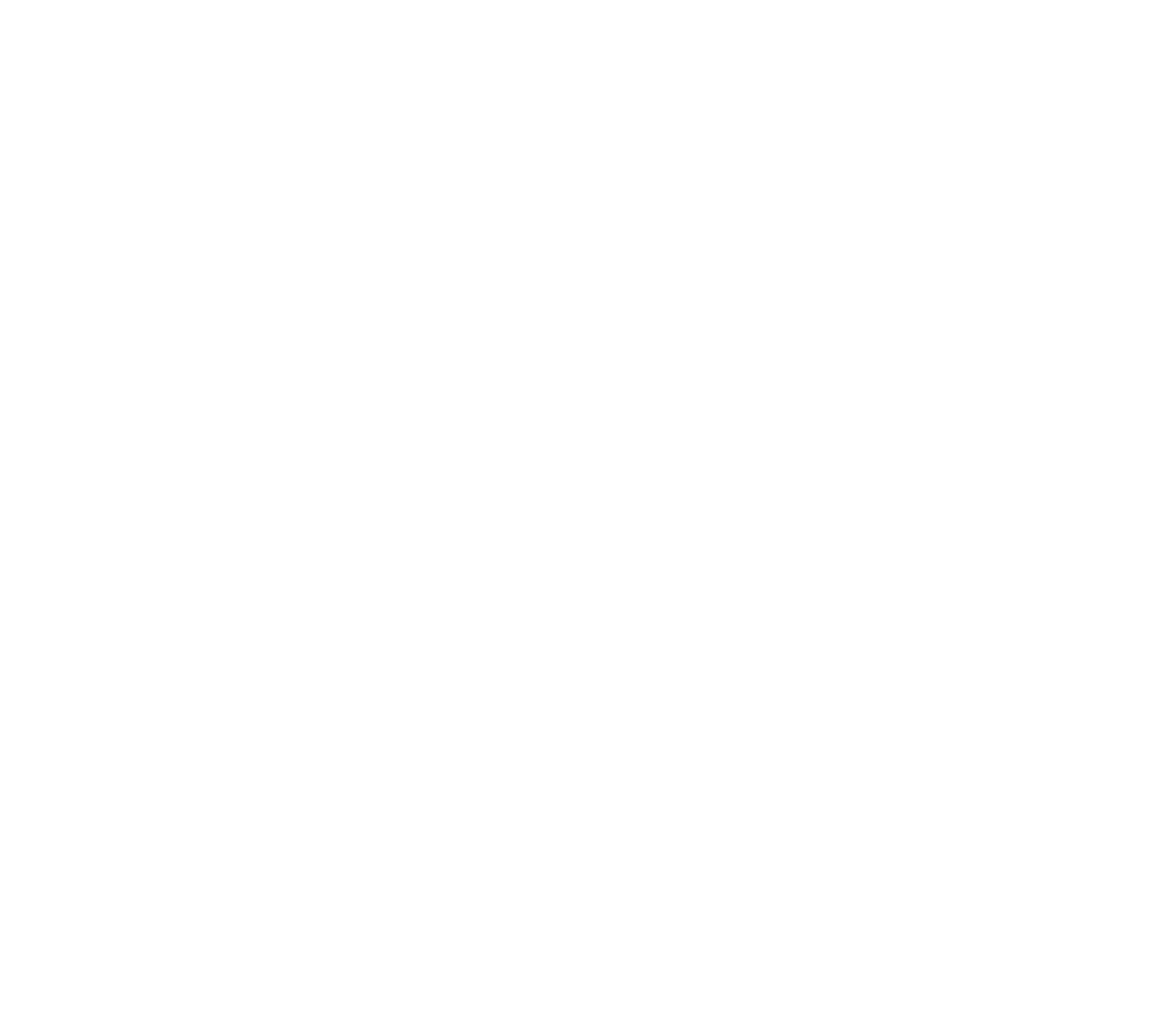}
    \caption{\label{poudre_microscope} (a) SEM image of the boehmite raw powder. The white circles highlight two representative powder agglomerates. (b)~TEM image of the boehmite dispersion in ethanol after sonication and drying. (c) and (d)~Cryo-TEM observation of boehmite crystallites, which consist in stacked sheets spaced by about 0.75~nm. Images obtained on a 4\% vol. boehmite gel diluted twice with distilled water. White bars indicate the scale in each picture, respectively (a) 100~$\mu$m, (b) and (c) 50~nm, and (d) 10~nm}
\end{figure}

\subsection{Characterizing boehmite powder and crystallites by SEM, TEM, and cryo-TEM}
\label{sec:electron}

\begin{figure}[t!]
    \centering
    \includegraphics[width=1\linewidth]{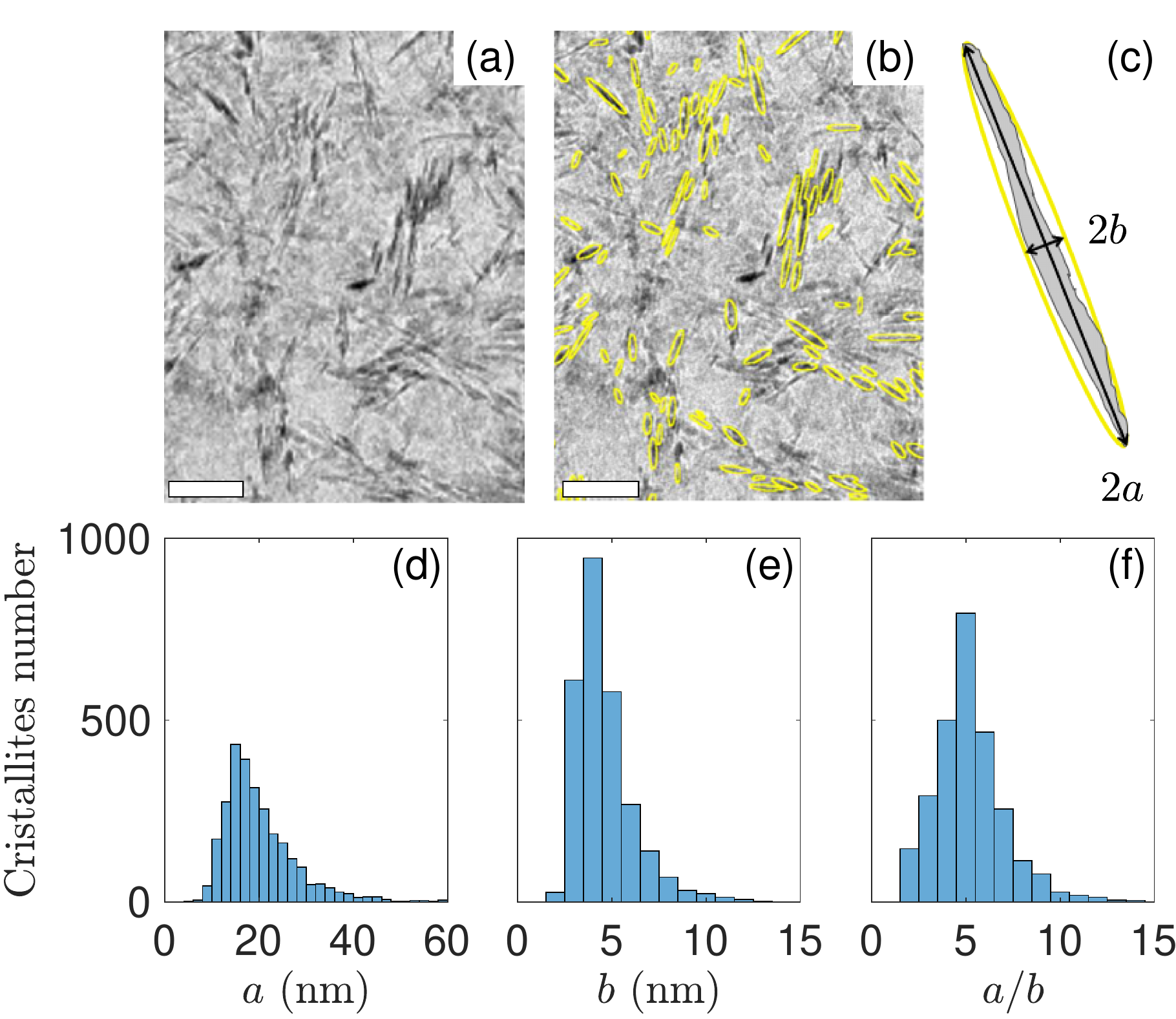}
    \caption{\label{fig:CRYO_TEM} (a) Cryo-TEM images of crystallites contained in a 4~\% vol. boehmite gel. (b)~Same image as in (a) where individual crystallites seen from the edge are circled by an ellipse with the same surface area, orientation, and centroid as the crystallite. White bars represent 50~nm. (c)~Sketch of a crystallite seen from the edge enclosed by its ellipse whose major (resp. minor) axis length is $a$ (resp. $b$). Probability density function (pdf) of (d) the major axis $a$, (e) the minor axis $b$, and (f) the aspect ratio $a/b$ of elliptical equivalents of the crystallites. Distributions obtained by analyzing 2712 crystallites identified in 5 independent images.}
\end{figure}

Figure~\ref{poudre_microscope}(a) shows that the agglomerates that constitute the boehmite powder used in this work have a diameter of a few tens of micrometers, in agreement with the median diameter of the raw boehmite powder of 45~$\mu$m reported by the manufacturer. These agglomerates are constituted of platelets referred to as \textit{crystallites}. These crystallites can be observed by TEM [Fig.~\ref{poudre_microscope}(b)] and cryo-TEM [Fig.~\ref{poudre_microscope}(c)]. The sheet-like microstructure of boehmite crystallites is particularly visible in cryo-TEM images [Fig.~\ref{poudre_microscope}(d)]. Five cryo-TEM images were further analyzed to characterize quantitatively the size distribution of the crystallites. In Fig.~\ref{fig:CRYO_TEM}(a), crystallites that appear dark are seen from the edge, i.e., from their thinner side, whereas crystallites that appear in lighter gray are seen from their large, flat side. The size analysis was performed on the crystallites that are seen from the edge using the FiJi software. These crystallites were identified by low-pass filtering the images in order to remove the background and keep only the elongated objects, which were then isolated by thresholding the images. Finally, the projected shape of the crystallites was assimilated to ellipses, whose main axis $a$ corresponds to the width of the platelet and the small axis $b$ to its thickness. The distributions of $a$ and $b$ as well as of the aspect ratio $a/b$ are shown in Figs.~\ref{fig:CRYO_TEM}(d)-(f) and yield $a=18\pm 8$~nm, $b=4\pm2$~nm, and $a/b=5\pm 2$. The large relative standard deviation of about 40~$\%$ indicates that the crystallites are rather polydisperse.

\subsection{Characterizing the size of ``unbreakable'' aggregates by DLS}
\label{sec:DLS}

\begin{figure}[t!]
    \centering
    \includegraphics[width=1\linewidth]{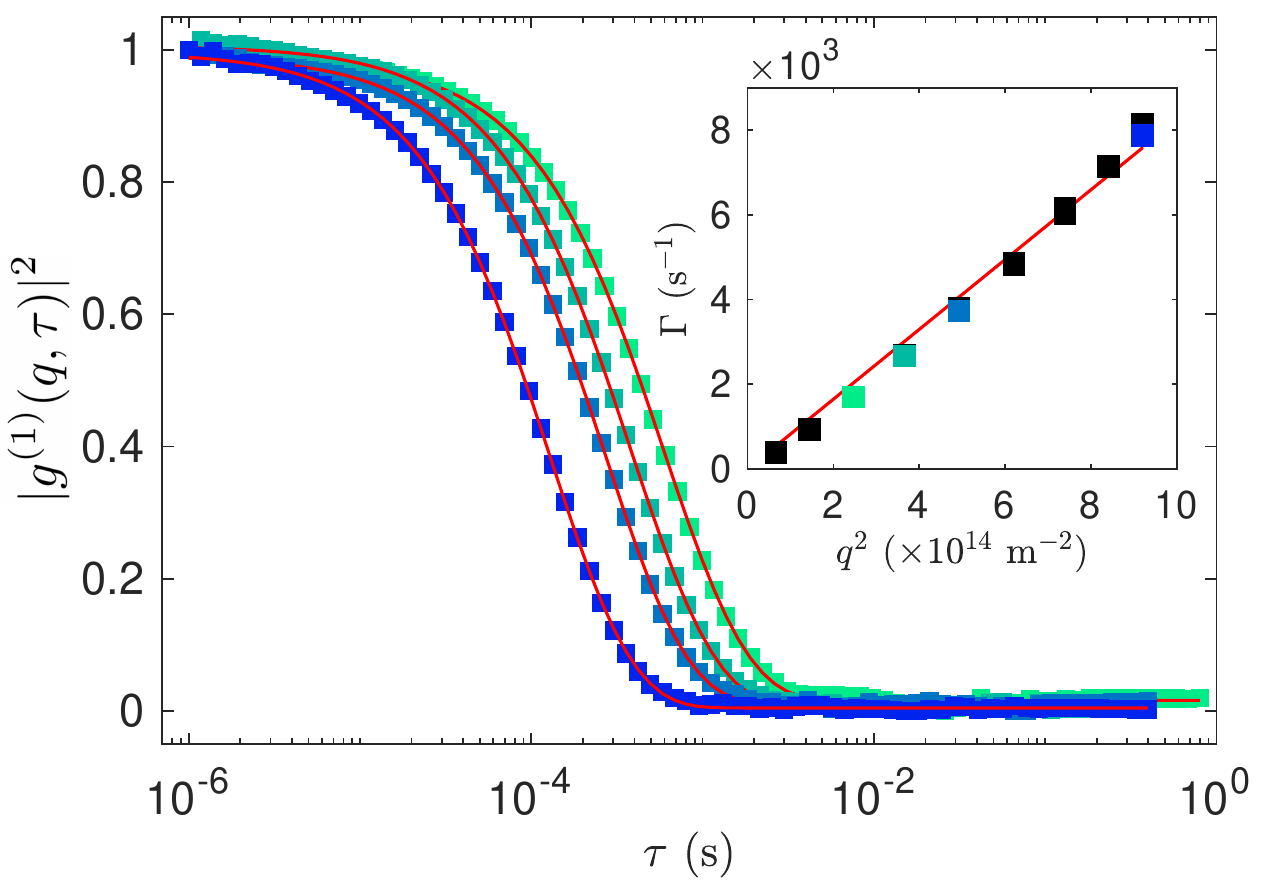}
    \caption{Squared modulus of the autocorrelation function $g^{(1)}(q,\tau)$ measured for $\theta= 60 \degree$ (\textcolor{colorb1}{$\blacksquare$}), $75 \degree$ (\textcolor{colorb2}{$\blacksquare$}), $90 \degree$ (\textcolor{colorb3}{$\blacksquare$}), and $150 \degree$~(\textcolor{colorb4}{$\blacksquare$}) on a primary gel diluted by a factor of 217 in distilled water with manual homogenization (see Table~\ref{tab:DLS_eau}). The red curves are the best fits of the experimental data using Eq.~\eqref{eq:fit_dls}. Inset: fit parameter $\Gamma$ vs.~$q^{2}$. The red line is the best linear fit of the data using Eq.~\eqref{eq:fit_gamma_dls_bis}, yielding a diffusion coefficient $D=4.1$~$\mu$m$^2$.s$^{-1}$.}
    \label{figure:Fit_DLS}
\end{figure}

Cryo-TEM images in Figs.~\ref{fig:CRYO_TEM}(a) and \ref{fig:CRYO_TEM}(b) suggest that the gel backbone is constituted of groups of crystallites fused together, which we shall hereafter refer to as \textit{``unbreakable'' aggregates}. In order to estimate the size of these aggregates, we have performed Dynamic Light Scattering (DLS) measurements on a 4~\% vol. boehmite gel at various degrees of dilution. 
After addition of distilled water, three different homogenization protocols were used: (i)~manual homogenization, (ii)~shear homogenization under a shear rate of 1000~s$^{-1}$ for 10~min in a Taylor-Couette cell of gap 1.0~mm connected to a stress-controlled rheometer, and (iii)~sonication for 1~min at 200~W using a probe sonicator (Hielscher UP200St). 

Figure~\ref{figure:Fit_DLS}(a) shows a selection of autocorrelation functions $\vert g^{(1)}(q,\tau)\vert^2$ obtained on a diluted primary gel for different scattering angles $\theta$, i.e., different values of $q$. In practice, we compute $\vert g^{(1)}(q,\tau)\vert^2$ from the measured $g^{(2)}(q,\tau)$ using $\vert g^{(1)}(q,\tau)\vert^2= \left(g^{(2)}(q,\tau) -\langle I(q,t)\rangle^2 \right)/ I_0(q)^2$, where $\langle I(q,t)\rangle^2$ is the measured baseline of $g^{(2)}(q, \tau)$, and $I_0(q)^2$ denotes its extrapolation to $\tau=0$ for each $q$ value. We model these results by assimilating an ``unbreakable'' aggregate to a spherical Brownian particle, which hydrodynamic radius $R_H$ obeys the Stokes-Einstein relation \cite{Einstein:1956}: $R_H = k_{B} T/(6 \pi \eta_{s} D)$, where $k_{B}$ is the Boltzmann constant, $D$ is the particle diffusion coefficient, and $\eta_{s}= 1$~mPa.s the solvent viscosity, here water. For a monodisperse suspension of such particles, $\vert g^{(1)}(q,\tau)\vert^2$ is expected to decay exponentially with the lag time $\tau$, with a characteristic time $\Gamma^{-1}$ such that \cite{Berne:2000}: 
\begin{equation}
\label{eq:fit_gamma_dls_bis}
    \Gamma(q) =  2 D q^{2}\,.
\end{equation}
Here, in order to take into account the large polydispersity of the ``unbreakable'' aggregates, we further assume that the aggregate size follows a Gaussian distribution characterized by a relative standard deviation $\sigma$. In the framework of the cumulant method introduced in Ref.~\cite{Koppel:1972}, we expect:
\begin{equation}
\label{eq:fit_dls}
    \vert g^{(1)}(q,\tau)\vert^2 = \exp\left(-\Gamma \tau \right)\left(1+\frac{\mu_{2}}{2}\tau^{2}\right) ,
\end{equation}
where $\mu_{2}=\sigma^2\Gamma^2$ and $\Gamma$ are used as adjustable parameters to fit the experimental data. As shown by the red curves in Fig.~\ref{figure:Fit_DLS}, Eq.~\eqref{eq:fit_dls} provides an excellent description of the data. The corresponding fit parameters are gathered in Table~\ref{tab:example_fit_DLS}. 
We find that the size distribution of this sample is characterized by a relative standard deviation $\sigma=0.4$, which corresponds to a broad-size polydispersity. This result is robustly observed for all the samples investigated, for which $\sigma$ ranges between 0.3 and 0.6. 

\begin{table}[t!]
\caption{\label{tab:example_fit_DLS} Values of the parameters used in Eq.~\eqref{eq:fit_dls} to fit the experimental curves of Fig. \ref{figure:Fit_DLS}.}
\centering
\begin{tabular}{p{1cm}p{1.5cm}p{1.5cm}p{1cm}}
\\
$\theta$ & $\Gamma$ (s$^{-1}$) & $\mu_{2}$ (s$^{-2}$) & $\sigma$  ($\%$)\\
\hline
\hline
\\
60$\degree$ & $1.5 \cdot 10^{3}$ & $3.7 \cdot 10^{5}$ & 41 \\
75$\degree$ & $2.5 \cdot 10^{3}$ & $1.2 \cdot 10^{6}$ & 44  \\
90$\degree$ & $3.5 \cdot 10^{3}$ & $1.6 \cdot 10^{6}$ & 36 \\
150$\degree$ & $7.5 \cdot 10^{3}$ & $9.2 \cdot 10^{6}$ & 40 \\
\end{tabular}
\end{table}

Moreover, the decay rate $\Gamma$ increases linearly with $q^{2}$  (see inset in Fig.~\ref{figure:Fit_DLS}), which justifies \textit{a posteriori} our assumptions, and allows us to estimate the diffusion coefficient $D$ and hence the mean hydrodynamic radius $R_H$ of the ``unbreakable'' aggregates. Table~\ref{tab:DLS_eau} reports the values of $R_H$ found for all 11 diluted gel samples. In the case of manual homogenization, $R_H$ significantly decreases from $(70 \pm 2)$~nm to $(52 \pm 2)$~nm with an increase of the dilution ratio. However, a dilution protocol that includes a high-shear homogenization period yields $R_H=(55 \pm 5)$~nm independent of the dilution ratio. This suggests that, for the smaller dilution ratios, the agglomerates were not completely broken down by manual homogenization. Finally, sonication leads to $R_H=(25 \pm 2)$~nm, which is twice as small as for the shear-homogenized dilutions. This suggests that the ``unbreakable'' aggregates are split in half by sonication. We conclude that the aggregates that constitute the microstructure of boehmite gels are characterized by a mean hydrodynamic radius $R_H=(55 \pm 5)$~nm and that they are ``unbreakable'' under shear, for shear rates lower than $\gp = 1000$~s$^{-1}$, which corresponds to the maximum shear rate used below in Sect.~\ref{sec:results}.

\subsection{Characterizing the structure of ``unbreakable'' aggregates by SAXS}

In order to characterize the ``unbreakable'' aggregates beyond their mean hydrodynamic radius, we used SAXS to investigate the microstructure of diluted samples of boehmite gels. In practice, the scattering patterns measured on such dilutions are isotropic, which allows us to focus on the azimuthal average of the scattered intensity over the whole range of polar angles $\varphi$ within the scattering plane for each $q$. Note, however, that isotropy only translates the lack of orientational order between the aggregates in the diluted dispersions, and does not imply that ``unbreakable'' aggregates are themselves isotropic.

\begin{figure}[t!]
    \centering
    \includegraphics[width=1\linewidth]{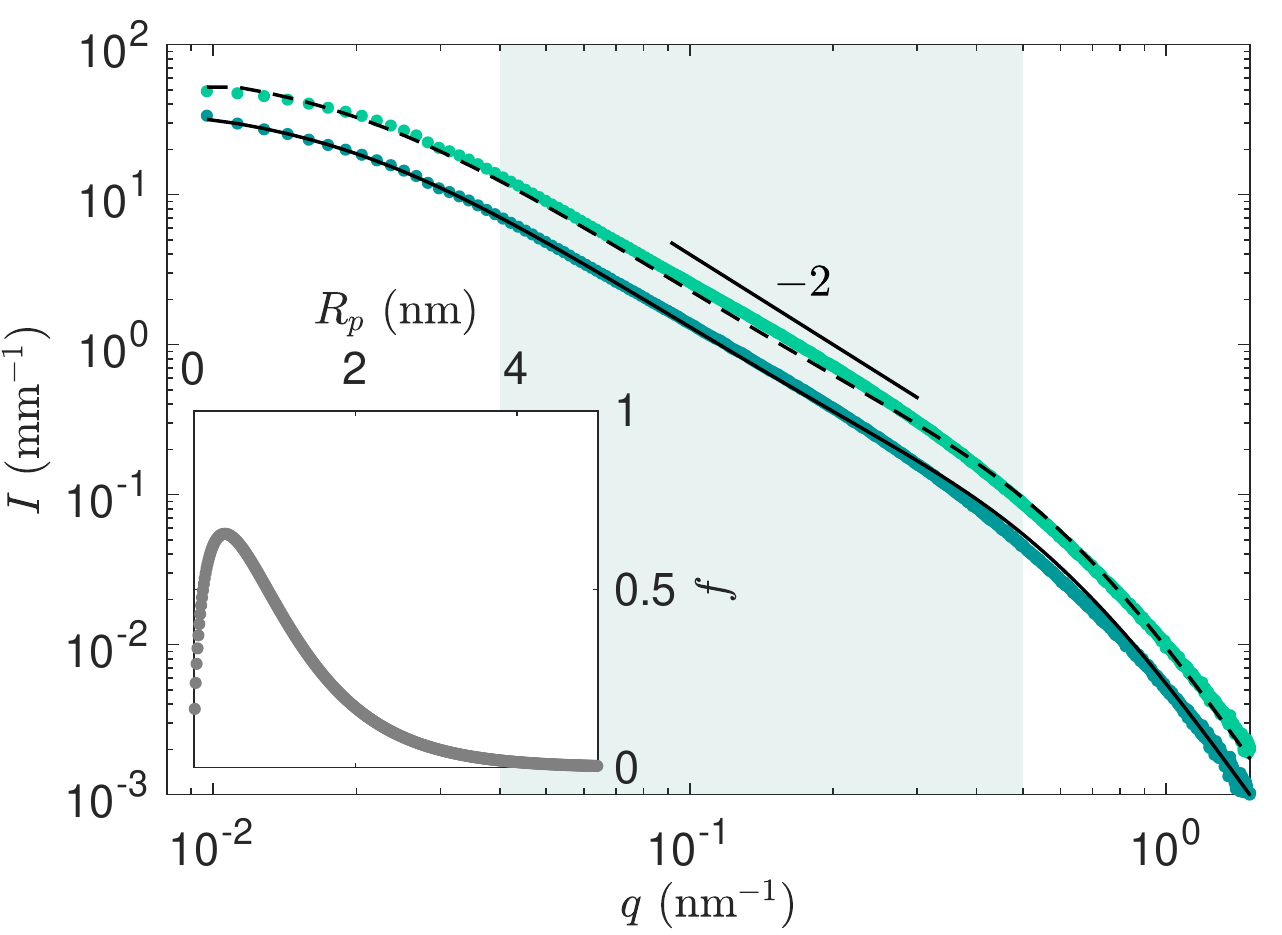}
    \caption{\label{fig:I_dilution_fit} Azimuthal average of the normalized and background-subtracted scattered intensity $I$ as a function of the amplitude $q$ of the scattering wave vector for a 4~\% vol. boehmite gel diluted by a factor of 51 (\textcolor{colorg1}{$\bullet$}) and 101 (\textcolor{colorg2}{$\bullet$}). The best fits of the data by  Eqs.~\eqref{eq:fit_dilution_SAXS}--\eqref{eq:S(q)} are shown in dashed and continuous lines, respectively. A power-law fit of $I(q)$ within the range $q \in [0.04,0.5]$~nm$^{-1}$ highlighted by the green shaded area yields an exponent $-2$. Inset: Schulz distribution $f$ used to model the size polydispersity through $R_{p}$ 
   in Eqs.~(\ref{eq:fit_dilution_SAXS})--(\ref{eq:S(q)}).}
\end{figure}

\begin{figure*}[t!]
    \centering
    \includegraphics[width=1\linewidth]{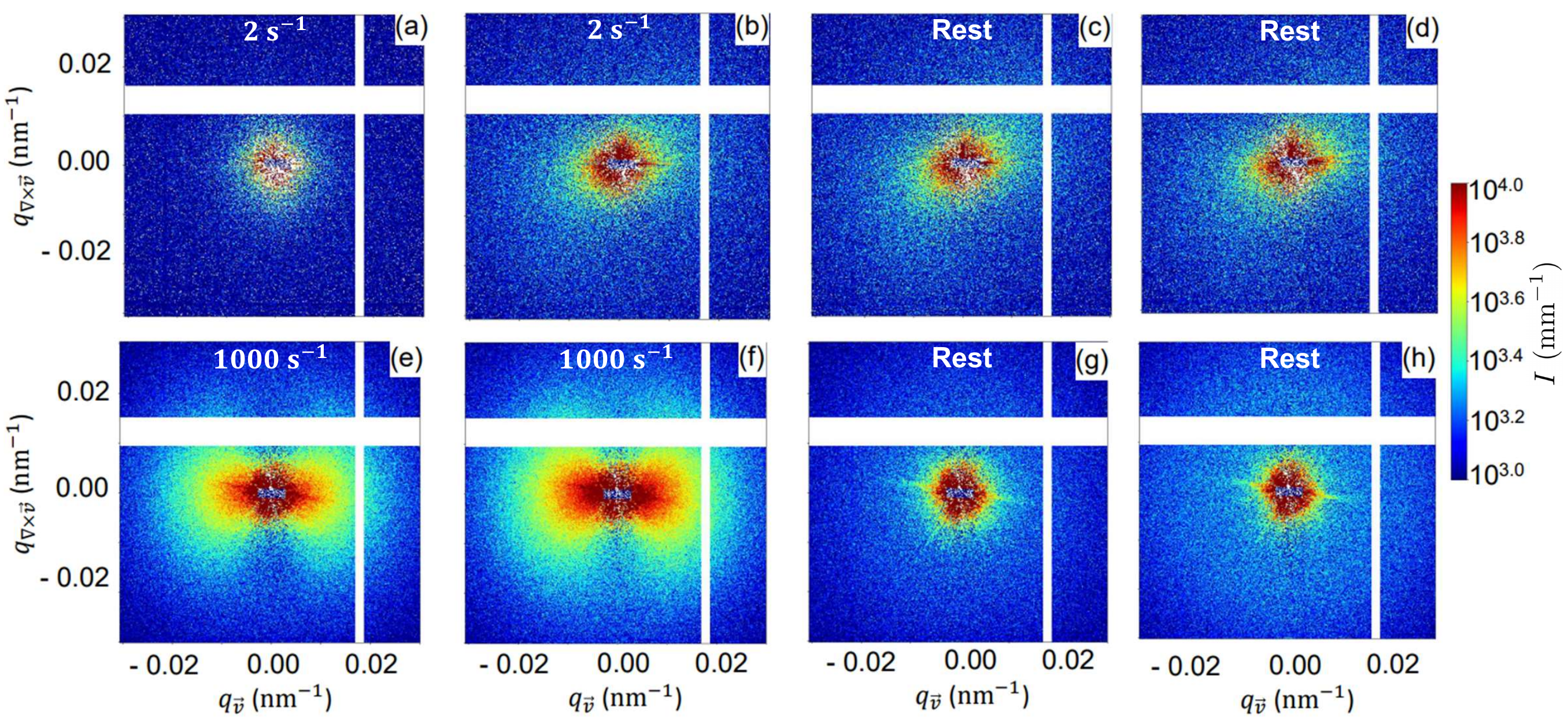}
    \caption{2D SAXS patterns measured on a 4~\% wt. boehmite gel in the rheo-SAXS setup with radial configuration: (a) first and (b) last scattering patterns measured during a 600~s shear step at $\gpp=2$~s$^{-1}$, (c) first and (d) last scattering patterns measured during a 600~s rest period following flow cessation. (e)--(h) Same measurements for $\gpp=1000$~s$^{-1}$. The scattered intensity value is coded according to the color scale on the right side of the figure. Scattering patterns were averaged over 2~s in (a)--(d) and over 3~s in (e)--(h). }
    \label{fig:spectre_2D_anisotropy}
\end{figure*}

Figure~\ref{fig:I_dilution_fit} displays the 1D scattering profile obtained for dilution ratios of 51 and 101. 
Whatever the dilution ratio, $I(q)$ shows a power-law decay with an exponent close to $-2$ for $0.04 \lesssim q \lesssim 0.5$~nm$^{-1}$, which hints at
fractal-like aggregates \cite{Schaefer:1984,Hyeon:1998,Rieker:2000}. 
We further model $I(q)$ as the intensity scattered by an assembly of crystallites of form factor $P(q)$ arranged into ``unbreakable'' aggregates with a structure factor $S(q)$, i.e.,  
\begin{equation}
\label{eq:fit_dilution_SAXS}
    I(q) \propto \phi V_{p} P(q)S(q)\,,
\end{equation}
where $\phi$ is the boehmite volume fraction within the dilution and $V_{p}$ the volume of an elementary crystallite. 
For the sake of simplicity, we assimilate the crystallites to spherical particles of radius $R_p$, so that $V_p=4/3 \pi R_{p}^3$. Under this assumption,  $P(q)$ and $S(q)$ can be expressed as \cite{Mildner:1986,Teixeira.1988}:
\begin{equation}
     P(q)=F(qR_{p})^2\,,
\end{equation}
where $F(x)=3(\sin x - x \cos x)/x^3$, and
\begin{equation}
\label{eq:S(q)}
    S(q)=1+\frac{d_{f}\Gamma(d_{f}-1)}{[1+1/(q^2\xi^2)]^{(d_{f}-1)/2}} \frac{\sin[(d_{f}-1)\arctan(q\xi)]}{(qR_{0})^{d_{f}}} ,
\end{equation}
where $d_{f}$ is the mass fractal dimension of the ``unbreakable'' aggregates, $\xi$ is the maximum size of the aggregates, i.e., the cutoff length above which the mass distribution is no longer described by the fractal model, $R_{0}$ is the mean radius of the crystallites, and $\Gamma$ denotes the Gamma function $\Gamma(z)= \int_{0}^{+\infty} t^{z} \text{e}^{-t} \text{d}t$. Finally, we take into account the  polydispersity of the crystallites by averaging Eqs.~(\ref{eq:fit_dilution_SAXS})--(\ref{eq:S(q)}) over $R_p$ following
a Schulz distribution with a standard deviation of $0.8$~nm (see inset in Fig.~\ref{fig:I_dilution_fit}) \cite{Kotlarchyk.1988}.

The data reported in Fig.~\ref{fig:I_dilution_fit} were fitted by Eqs.~(\ref{eq:fit_dilution_SAXS})--(\ref{eq:S(q)}) using the SasView software with $d_{f}$, $\xi$, and $R_0$ as free parameters \cite{SASview}.
The best fits of the experimental data yielded 
$d_{f}=2.05 \pm 0.05$, $\xi= (53 \pm 5)$~nm, and $R_{0}=(1.1 \pm 0.1)$~nm, for both dilution factors.
The value for the size $\xi$ of the aggregates is in excellent agreement with the hydrodynamic radius $R_H$ inferred from DLS in Sect.~\ref{sec:DLS}. However, the average size $R_0$ of the crystallites is significantly smaller than the dimensions extracted from cryo-TEM images in Sect.~\ref{sec:electron}. This discrepancy most likely originates from the crude assumption that the crystallites are spherical, and from the specific shape of the distribution function accounting for the broad size polydispersity of the crystallites (see inset in Fig.~\ref{fig:I_dilution_fit}). Yet, the description of the building blocks in boehmite gels as  ``unbreakable'' fractal-like aggregates of crystallites, with typical size $\xi= (53 \pm 5)$~nm, offers a consistent description of the scattering data. When the boehmite dispersion is left at rest, these colloidal ``unbreakable'' aggregates experience attractive interactions and themselves assemble into a gel, i.e., a percolating network that provides some elasticity to the sample. 

\section{Evidence for shear-induced anisotropy in boehmite gels}
\label{sec:results}

Having characterized the structure of the building blocks in boehmite gels, we now turn to the impact of shear history upon their microstructure and rheology. More precisely, we perform time-resolved rheo-SAXS experiments to elucidate the impact of shear at a rate $\gpp$ both under flow and upon flow cessation.

\subsection{Time-resolved rheo-SAXS measurements}

A 4\% wt.~boehmite gel was rejuvenated at $\gp=1000$~s$^{-1}$, before being sheared at a rate of interest $\gpp$ for 600~s. Finally, the flow was stopped and the gel rebuilding was monitored over 600~s. This corresponds to protocol ``A'' introduced above in Sect.~\ref{sec:summary} yet with a shorter rest period. Figure~\ref{fig:spectre_2D_anisotropy} gathers a selection of rheo-SAXS patterns measured both under shear at $\gpp$ and after flow cessation, for $\gpp=2$~s$^{-1}$ [Fig.~\ref{fig:spectre_2D_anisotropy}(a)--(d)] and $\gpp=1000$~s$^{-1}$ [Fig.~\ref{fig:spectre_2D_anisotropy}(e)--(h)].
At the start of the shear step at $\gpp=2$~s$^{-1}$, the scattering pattern appears to be isotropic [Fig.~\ref{fig:spectre_2D_anisotropy}(a)] and becomes progressively anisotropic under shear, i.e., the scattered intensity depends on the polar angle~$\varphi$ within the scattering plane (velocity $\vec{v}$, vorticity~$\nabla\times \vec{v}$) [Fig.~\ref{fig:spectre_2D_anisotropy}(b)]. More precisely, following a shear period of 600~s  at $\gpp=2$~s$^{-1}$, the scattered intensity for wavenumbers $q> 0.01$~nm$^{-1}$ is larger along the velocity direction $\vec{v}$ than along the vorticity direction $\nabla \times \vec{v}$. Moreover, such anisotropy of the scattering pattern persists upon flow cessation, as measured right after flow cessation in Fig.~\ref{fig:spectre_2D_anisotropy}(c), as well as 600~s later in Fig.~\ref{fig:spectre_2D_anisotropy}(d). This observation points towards the growth of some anisotropy within the gel microstructure under application of a low shear rate $\gpp$. Since the scattering pattern does not evolve significantly after shear at $\gpp=2$~s$^{-1}$ is stopped, we conclude that this anisotropy remains frozen upon flow cessation.

Remarkably, the scenario reported in Fig.~\ref{fig:spectre_2D_anisotropy}(e)--(h) for the same gel sheared at $\gpp=1000$~s$^{-1}$ is very different. First, when such a large shear rate is applied, the scattering pattern becomes strongly anisotropic almost instantaneously, and displays a ``butterfly'' shape oriented along the velocity direction [Fig.~\ref{fig:spectre_2D_anisotropy}(e)], which becomes even more pronounced over 600~s [Fig.~\ref{fig:spectre_2D_anisotropy}(f)]. 
Such a ``butterfly'' pattern has been reported from scattering measurements in other colloidal dispersions such as aqueous laponite gels \cite{Pignon:1997}, thermoreversible silica gels 
\cite{Varadan:2001,Narayanan.2020,Hoekstra:2005}, silica-filled rubbers \cite{Staropoli:2019}, and carbon-black-filled elastomers \cite{Beutier:2022}, or from confocal microscopy in flocculated suspensions of PMMA spheres \cite{Massaro:2020}. It has been linked to long-range hydrodynamic interactions, which lead to some ordering of the particles along the vorticity direction or even stabilize clusters of particles into vorticity-aligned flocs \cite{Varga:2018}. Although the ``butterfly'' pattern is of potential interest to better understand the microstructure of boehmite dispersions under a strong shear flow, the present work is rather concerned with the spectra observed after flow cessation. Strikingly, when shear at $\gpp=1000$~s$^{-1}$ is stopped, the scattering pattern abruptly loses its anisotropic shape [Fig.~\ref{fig:spectre_2D_anisotropy}(g)], and remains isotropic for the subsequent 600~s [Fig.~\ref{fig:spectre_2D_anisotropy}(h)].

\begin{figure}[t!]
    \centering
    \includegraphics[width=1\linewidth]{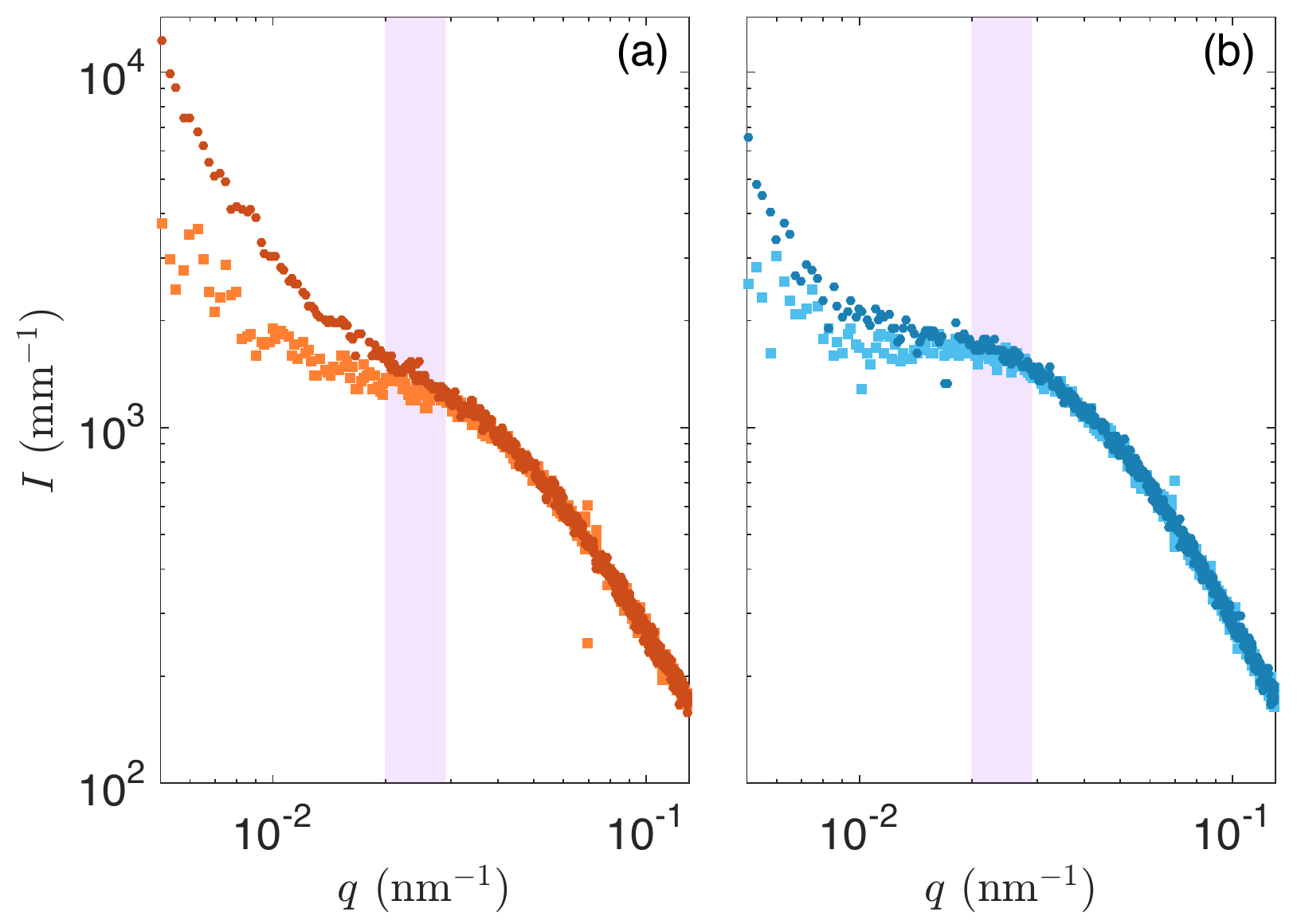}
    \caption{Intensity $I$ as a function of the amplitude $q$ of the scattering wave vector scattered along the velocity direction (darker $\bullet$  symbols obtained by averaging $I(q,\varphi)$ over $\varphi\in [-15^\circ,15^\circ]$) and along the vorticity direction (lighter $\blacksquare$ symbols obtained by averaging $I(q,\varphi)$ over $\varphi\in [75^\circ,105^\circ]$) by a 4~\% wt. boehmite gel probed 600~s after flow cessation following a shear period of 600~s at (a) $\gpp=2$~s$^{-1}$ and (b) $\gpp=1000$~s$^{-1}$. The purple area indicates the $q$-range used for averaging in Fig.~\ref{fig:fit_parameters_anisotropy}.}
    \label{fig:I_gel_primaire_2D_tailles}
\end{figure}

\begin{figure*}[t!]
    \centering
    \includegraphics[width=0.9\linewidth]{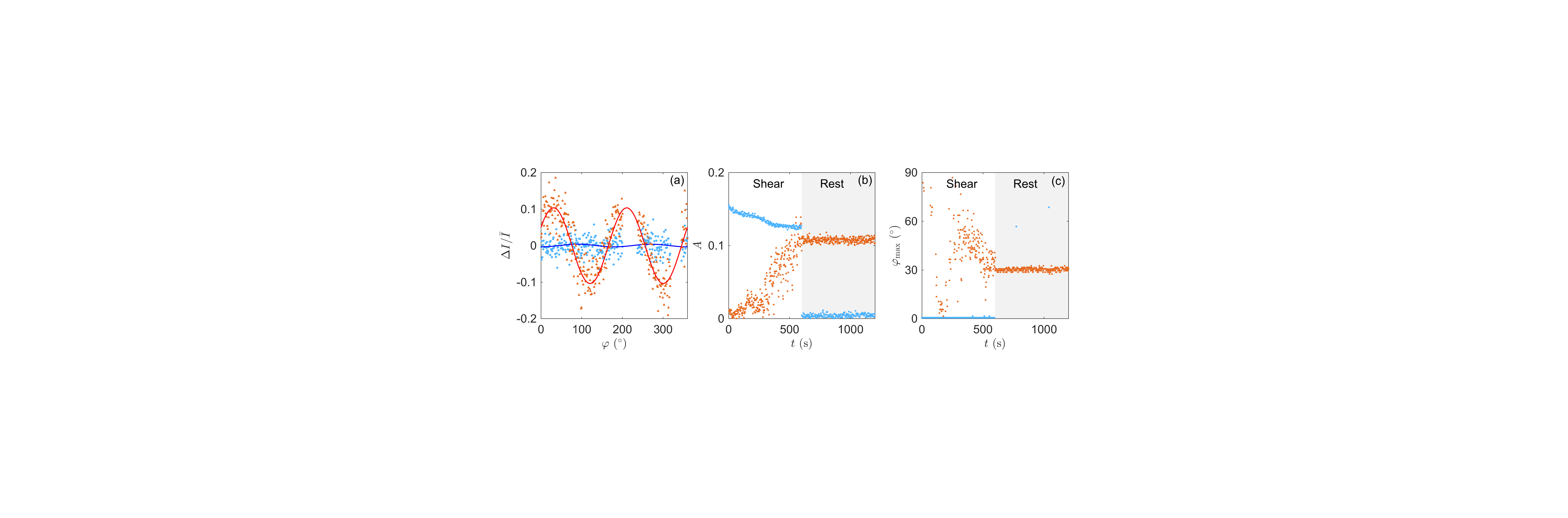}
    \caption{(a) $\Delta I / \bar{I}$ (see text for definition) as a function of the polar angle $\varphi$. Data recorded at $600$~s after cessation of shear at $\gpp=2$~s$^{-1}$ (\textcolor{orange}{$\blacktriangle$}) and $\gpp=1000$~s$^{-1}$ (\textcolor{cyan}{$\bullet$}). Solid lines are the best fits of the data to Eq.~(\ref{eq:fit_sin_SAXS}).
    The lack of data points for $\varphi\in [200^\circ,230^\circ]$ and $\varphi\in[315^\circ,340^\circ]$ is due to the gaps between the four sensors of the detector (see white stripes in Fig.~\ref{fig:spectre_2D_anisotropy}). Time evolution of (b)~the degree of anisotropy $A$ and (c)~the ``preferred'' orientation $\varphi_{\max}$ for $\gpp=2$~s$^{-1}$ (\textcolor{orange}{$\blacktriangle$}) and $\gpp=1000$~s$^{-1}$ (\textcolor{cyan}{$\bullet$}). The white background from $t=0$ to $t=600$~s identifies the shearing step at $\gpp$, while the gray background from $t=600$~s to $t=1200$~s corresponds to the rest period after flow cessation at $t=600$~s.
    }
\label{fig:fit_parameters_anisotropy}
\end{figure*}

In order to quantify the changes imparted by $\gpp$ to the boehmite gel microstructure over the most relevant range of length scales, Fig.~\ref{fig:I_gel_primaire_2D_tailles} compares the scattered intensity $I(q)$ computed along the flow direction by averaging the scattering pattern over the angular sector $\varphi\in [-15^\circ,15^\circ]$ to that computed along the vorticity direction by averaging over $\varphi\in [75^\circ,105^\circ]$. 
Applying a shear rate $\gpp=2$~s$^{-1}$ prior to flow cessation leads to strong differences between the averaged intensity along the flow direction and along the vorticity direction for $q \lesssim 0.03$~nm$^{-1}$, i.e., for length scales larger than about 200~nm [Fig.~\ref{fig:I_gel_primaire_2D_tailles}(a)]. In contrast, $\gpp=1000$~s$^{-1}$ leads to negligible differences over the whole accessible range of $q$ values [Fig.~\ref{fig:I_gel_primaire_2D_tailles}(b)].
In the following, we choose to focus on the range $q\in[0.020,0.029]$~nm$^{-1}$, which corresponds to length scales between 200~nm and 300~nm, i.e., 4 to 6 times the size of the ``unbreakable'' building blocks within the gel. As discussed in the next section, such a choice does not affect our general conclusions. Finally, for $q\gtrsim 0.06$~nm$^{-1}$, i.e., for length scales smaller than about 100~nm, all averaged spectra coincide with those of the dilutions from Fig.~\ref{fig:I_dilution_fit} once normalized by the volume fraction $\phi$. This confirms that the structure of the ``unbreakable'' aggregates is conserved throughout the rheological protocol.

\subsection{Quantitative analysis of shear-induced anisotropy}

As explained above, we consider the scattered intensity averaged over $q\in[0.020,0.029]$~nm$^{-1}$, denoted as  $I(\varphi,t)$ in the following to emphasize its dependence on time and on the polar angle $\varphi$. We further focus on the relative deviation of the scattered intensity from its angular average, $\Delta I / \bar{I}$, defined as:
\begin{equation}
    \frac{\Delta I}{\bar{I}}(\varphi,t) =  \frac{I(\varphi,t)-\bar{I}(t)}{\bar{I}(t)} ~~ \text{with} ~~ \bar{I}(t) =\frac{1}{2\pi} \int_{\varphi=0}^{2\pi} I(\varphi,t)  \textrm{d}\varphi\,.
    \label{eq:deltaI}
\end{equation}
Figure~\ref{fig:fit_parameters_anisotropy}(a) compares the angular dependence of $\Delta I / \bar{I}$ measured 600~s after flow cessation, either from $\gpp=2$~s$^{-1}$ or from $\gpp=1000$~s$^{-1}$.
The data for $\gpp=2$~s$^{-1}$ clearly reveal a sinusoidal dependence with $\varphi$ [see orange symbols in Fig.~\ref{fig:fit_parameters_anisotropy}(a)], which confirms quantitatively the anisotropy detected in Fig.~\ref{fig:spectre_2D_anisotropy}(d). For $\gpp=1000$~s$^{-1}$, however, $\Delta I / \bar{I}$ remains essentially constant and equal to 0 up to experimental uncertainty [see blue symbols in Fig.~\ref{fig:fit_parameters_anisotropy}(a)], consistently with the isotropic scattering reported in Fig.~\ref{fig:spectre_2D_anisotropy}(h).

We further quantify microstructural anisotropy by fitting $\Delta I / \bar{I}(\varphi,t)$ at all times $t$ using:
\begin{equation}
I_{\text{fit}}(\varphi,t) = A(t) \cos\left[ 2(\varphi-\varphi_{\max}(t))\right] \,,
\label{eq:fit_sin_SAXS}    
\end{equation}
where the amplitude $A(t)$ and the phase $\varphi_{\max}(t)$ are two adjustable parameters, which respectively yield the temporal evolution of the degree of anisotropy within the scattering pattern, and that of the orientation where anisotropy is maximal. In that framework, anisotropy is considered to be  significant when $A(t)>0.01$. 
Both these parameters are reported as a function of time in Figs.~\ref{fig:fit_parameters_anisotropy}(b) and \ref{fig:fit_parameters_anisotropy}(c) respectively. 
In agreement with the qualitative observations drawn from Figs.~\ref{fig:spectre_2D_anisotropy}(a)--(d), the degree of anisotropy slowly grows under application of shear at $\gpp=2$~s$^{-1}$, and  remains constant at $A(t)\simeq 0.1$ once shear is stopped at $t=600$~s [see orange symbols in Fig.~\ref{fig:fit_parameters_anisotropy}(b)]. Concomitantly, the ``preferred'' orientation $\varphi_{\max}(t)$ strongly fluctuates upon application of shear before decreasing progressively from about 60$\degree$ to 30$\degree$. Upon flow cessation, the value $\varphi_{\max}(t)\simeq 30\degree$ reached under shear is preserved during the whole rest period [see orange symbols in Fig.~\ref{fig:fit_parameters_anisotropy}(c)].
This unambiguously demonstrates that shear at $\gpp=2$~s$^{-1}$ induces some anisotropy in the gel microstructure at the scale of 200--300~nm that remains trapped upon flow cessation. More precisely, this also suggests that, at least for the range of length scales involved in our analysis, shear at $\gpp=2$~s$^{-1}$ leads to the formation of anisotropic groups of ``unbreakable'' aggregates with an average orientation of about 30$\degree$ relative to the velocity direction and that persist after shear is stopped.

\begin{figure*}[tb]
    \centering    \includegraphics[width=0.65\linewidth]{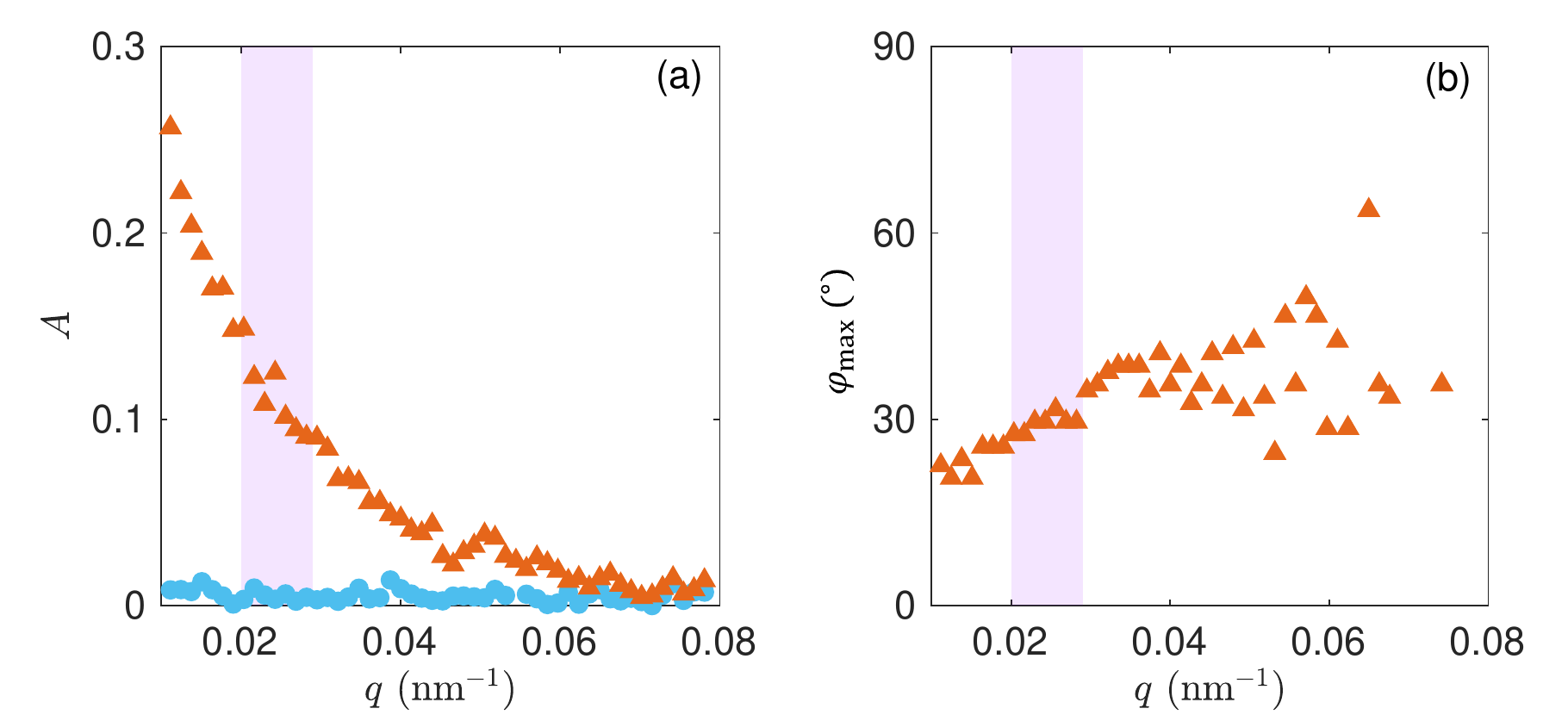}
    \caption{(a) Degree of anisotropy $A$ and (b)~``preferred'' orientation $\varphi_{\max}$ as a function of the amplitude $q$ of the scattering wave vector. Measurements performed 600~s after flow cessation
    following 600~s of shear at $\gpp=2$~s$^{-1}$ (\textcolor{orange}{$\blacktriangle$}) and at $\gpp=1000$~s$^{-1}$~(\textcolor{cyan}{$\bullet$}). The angle $\varphi_{\max}$ remains undetermined when $A<0.01$. The purple area indicates the $q$-range used for averaging in Fig.~\ref{fig:fit_parameters_anisotropy}.}
    \label{fig:fit_parameters_anisotropy_q}
\end{figure*}

In stark contrast with the previous results, the application of shear at $\gpp=1000$~s$^{-1}$ immediately induces a strong anisotropy oriented along the velocity direction ($A\simeq 0.1$--0.15 and $\varphi_{\max}=0\degree$) that vanishes almost instantaneously upon flow cessation ($A\simeq 0$ and $\varphi_{\max}$ cannot be measured) [see blue symbols in  Fig.~\ref{fig:fit_parameters_anisotropy}(b,c)]. This corresponds to the presence of a ``butterfly'' pattern that suddenly gives way to an isotropic pattern at $t=600$~s, as reported in Figs.~\ref{fig:spectre_2D_anisotropy}(e)--(h). 
We interpret this sequence of events as the  signature of the full break-up of the gel microstructure by shear at $\gpp=1000$~s$^{-1}$ into a dispersion of individual ``unbreakable'' aggregates whose positions show significant correlations along the vorticity direction. When shear is stopped, ``unbreakable'' aggregates reconnect in an isotropic fashion, i.e., they form a network of isotropic clusters without any longe-range order nor any orientational correlations. 

\begin{figure*}[t!]
    \centering
    \includegraphics[width=0.9\linewidth]{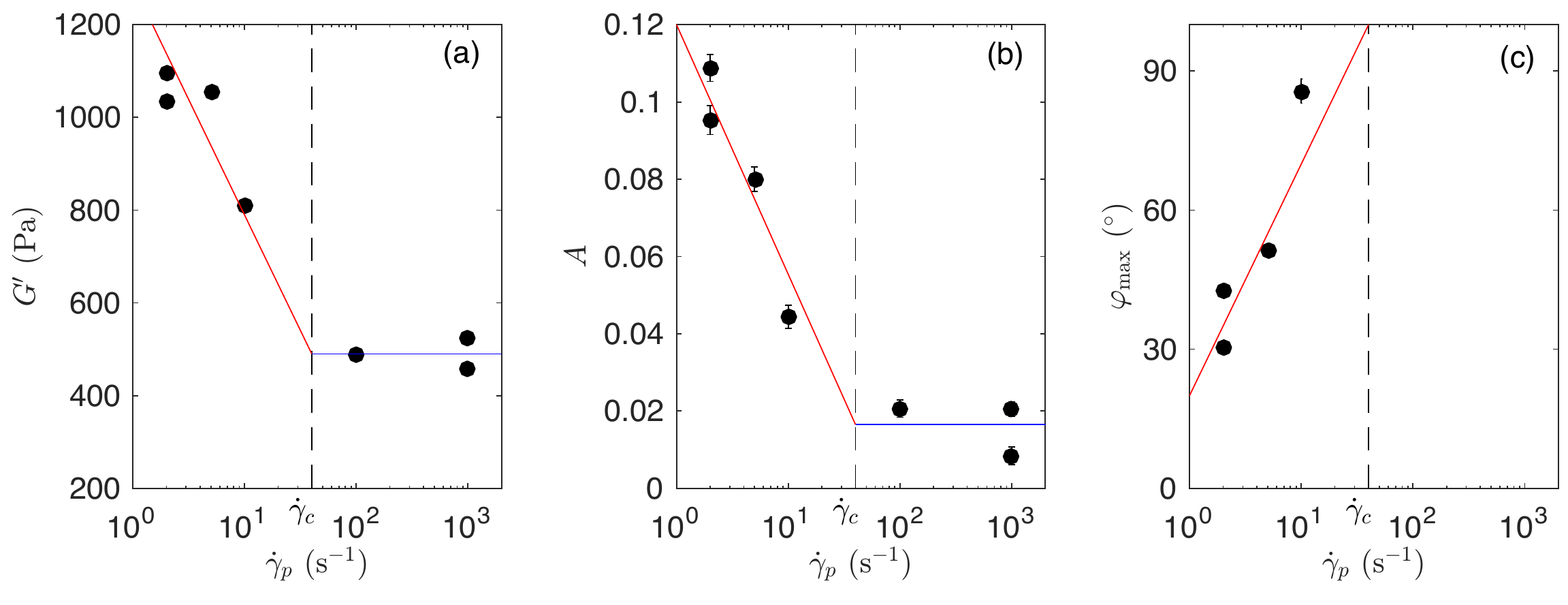}
    \caption{(a) Elastic modulus $G'$, (b)~degree of anisotropy $A$, and (c)~``preferred'' orientation $\varphi_{\max}$ as a function of the shear rate $\dot \gamma_p$ applied prior to flow cessation. Measurements performed 600~s after flow cessation and averaged over 20~s. Error bars represent the standard deviation and solids lines are guides to the eye. 
    }\label{fig:fit_parameters_anisotropy_gamma}
\end{figure*}

In order to check for the presence of anisotropy over the whole range of accessible length scales, we also estimated the degree of anisotropy $A$ and the ``preferred'' orientation angle $\varphi_{\rm max}$ for all available $q$-values. To this aim, we considered $I(q,\varphi)$ recorded at time $t=600$~s after flow cessation and averaged over $\pm 2$ data points around each value of $q$, i.e., over a narrow range $\delta q=0.0013$~nm$^{-1}$. For the various values of $q$, $\Delta I/\bar{I}$ was then computed and fitted to a sinusoid based on Eqs.~\eqref{eq:deltaI} and \eqref{eq:fit_sin_SAXS}, where $t$ should simply be replaced by $q$.
Figure~\ref{fig:fit_parameters_anisotropy_q} shows the fitting parameters $A$ and $\varphi_{\rm max}$ as a function of $q$. Following application of shear at $\gpp=2$~s$^{-1}$, the microstructure is significantly anisotropic for $q\lesssim 0.06$~nm$^{-1}$, i.e., for all length scales larger than the typical size of ``unbreakable'' aggregates [see orange symbols in Fig.~\ref{fig:fit_parameters_anisotropy_q}(a)]. Moreover, the degree of anisotropy increases from the scale of ``unbreakable'' aggregates up to the largest accessible length scale, here about 600~nm. On the other hand, after application of shear at $\gpp=1000$~s$^{-1}$, the gel displays an isotropic microstructure ($A\lesssim 0.01$) whatever the length scale under scrutiny [see blue symbols in Fig.~\ref{fig:fit_parameters_anisotropy_q}(a)]. Finally, Fig.~\ref{fig:fit_parameters_anisotropy_q}(b) shows that the ``preferred'' orientation in the scattering pattern measured 600~s after cessation of shear at $\gpp=2$~s$^{-1}$ depends only weakly on $q$, and shows an increase from about 25$\degree$ to about 45$\degree$ as $q$ spans the whole accessible range.

Finally, the above rheo-SAXS experiments were repeated for various shear rates $\gpp$. The results are gathered in Fig.~\ref{fig:fit_parameters_anisotropy_gamma}.
As expected from Refs.~\cite{Sudreau:2022a,Sudreau:2022b}, the elastic modulus $G'$ measured after the 600~s rest period following application of shear at $\gpp$ displays the same trend as in Fig.~\ref{fig:robustesse_protocole}: above the critical shear rate $\gpc\simeq 40$~s$^{-1}$, $G'$ is insensitive to $\gpp$, whereas it strongly increases when $\gpp$ decreases below $\gpc$ [Fig.~\ref{fig:fit_parameters_anisotropy_gamma}(a)]. Remarkably, the degree of anisotropy achieved 600~s after flow cessation follows the very same evolution with $\gpp$ as the elastic modulus [Fig.~\ref{fig:fit_parameters_anisotropy_gamma}(b)]. This demonstrates that the reinforcement of elasticity below $\gpc$ is linked to flow-induced microstructural anisotropy. Furthermore, the ``preferred'' orientation angle $\varphi_{\max}$ is observed to decrease sharply from about 90~$\degree$ to about 30~$\degree$ as $\gpp$ decreases below $\gpc$ [Fig.~\ref{fig:fit_parameters_anisotropy_gamma}(c)]. This indicates that the average orientation of clusters of ``unbreakable'' aggregates turns from the vorticity to the velocity direction as clusters are generated under lower shear rates.

\section{Discussion and conclusion}
\label{sec:discussion}

With the present work, we have shown that the building blocks of boehmite gels are composed of ``unbreakable'' fractal-like aggregates of anisotropic colloidal crystallites. Due to acid-induced attractive interactions, these aggregates, whose size is about 50~nm, form a percolated network that displays a pronounced sensitivity to shear: when exposed to shear rates $\gpp$ below a critical value $\gpc$, their elasticity is significantly enhanced. 
The present findings comfort the interpretation first proposed in Ref.~\cite{Sudreau:2022a}, where shear at $\gpp>\gpc$ is sufficient to fully disperse the aggregates and, as such, rejuvenate the sample microstructure that reforms anew upon flow cessation. In contrast, shear at $\gpp<\gpc$ does not fully break the gel microstructure, which therefore bears some memory of the flow, once the flow is stopped. Moreover, the lower the shear rate applied before flow cessation, the stronger the elastic properties of boehmite gels measured after flow cessation. Based on rheology alone, we have referred to this phenomenon as ``overaging,'' considering the sensitivity of such reinforcement to the total strain accumulated before flow cessation \cite{Sudreau:2022b}.

Here, based on rheo-SAXS measurements, we have shown that such a reinforcement of the viscoelastic properties is concomitant with a change in the sample microstructure, which becomes anisotropic at length scales larger than about 5 times the size of the ``unbreakable'' aggregates composing the gel. For low shear rate, i.e., for $\gpp<\gpc$, the anisotropy of the sample microstructure grows under shear and levels off,  upon flow cessation, to a constant value, which increases for decreasing shear rates. This provides strong evidence for a deep correlation between shear-induced reinforcement and the growth of anisotropy within the gel microstructure. 

Furthermore, our results can be interpreted in light of recent experiments performed on dispersions of anisotropic colloidal particles of the same size as the present ``unbreakable'' aggregates but with increasing aspect ratio. A direct comparison of their viscoelastic properties at fixed volume fraction demonstrates that anisotropic building blocks yield stronger elastic properties \cite{Kao:2022}. In our case, clusters of ``unbreakable'' aggregates play the role of anisotropic building blocks, which may account for the weaker reinforcement reported here compared to that observed in Ref.~\cite{Kao:2022}. In that framework, our results suggest that moderate shear can be used to imprint some anisotropy into the microstructure of a colloidal gel and, therefore, tune its mechanical properties. Future measurements and analyses should focus on investigating the architecture of boehmite at scales larger than a few times the size of ``unbreakable'' aggregates, in order to fully unveil the hierarchical nature of shear-induced anisotropy. Beyond boehmite gels, it also remains to be seen how generic this low-shear reinforcement scenario is, and what are the minimal ingredients to make it happen, since depletion gels show the opposite behavior \cite{Koumakis:2015}. Our results call for additional experiments probing the impact of shear history on dispersions of colloids with large aspect ratios and attractive interparticle forces from different origins.  

\begin{acknowledgements}
The authors thank V.~Dolique for his help with the SEM observations, and E.~Del Gado, O.~Diat, I.~H{\'e}naut, M.~Lattuada, E.~L{\'e}colier,  G.H.~McKinley, M.~Meireles, M.~Morbidelli, T.~Narayanan, G.~Petekidis, and A.~Poulesquen for fruitful discussions. ESRF is acknowledged for the provision of synchrotron beamtime (SC-5113).
\end{acknowledgements}

\section*{References}
%

 \end{document}